\documentclass[aps,prd,onecolumn,groupedaddress,showpacs,nofootinbib,amssymb]{revtex4}
\usepackage[dvips]{graphicx}
\usepackage{amssymb}
\usepackage{amsmath}
\usepackage{graphicx,,color}
\usepackage{amsfonts}
\usepackage{bm}
\usepackage{cancel}
\usepackage{comment}

\newcommand{\bea}{\begin{eqnarray}}
\newcommand{\eea}{  \end{eqnarray}}

\newcommand\be{\begin{equation}}
\newcommand\ee{\end{equation}}

\newcommand{\ba}{\begin{eqnarray}}
\newcommand{\ea}{\end{eqnarray}}
\allowdisplaybreaks[4]

\allowdisplaybreaks[4]

\begin{document}

\title{String Corrected Scalar Field Inflation Compatible with the ACT Data}
\author{V.K. Oikonomou,$^{1,2}$}\email{voikonomou@gapps.auth.gr;v.k.oikonomou1979@gmail.com}
\affiliation{$^{1)}$Department of Physics, Aristotle University of
Thessaloniki, Thessaloniki 54124, Greece\\
$^{2)}$ Center for Theoretical Physics, Khazar University, 41
Mehseti Str., Baku, AZ-1096, Azerbaijan}

\tolerance=5000

\begin{abstract}
We consider the impact of the first string corrections of
minimally coupled single scalar field theory on inflationary
dynamics. Specifically we consider separately the string
corrections $\sim \alpha'\xi(\phi)c_2\,\left( \partial_{\mu}\phi
\partial^{\mu}\phi\right)^2$ and $\sim \alpha'c \xi(\phi)\square
\phi
\partial_{\mu}\phi \partial^{\mu}\phi$, where $\alpha'$ is the square of the string scale. Our aim is to
develop a theory which is self consistent in the sense that the
field equations reproduce themselves in the slow-roll
approximation. Such a requirement for the theory with $\sim
\alpha'\xi(\phi) c_2\left(
\partial_{\mu}\phi \partial^{\mu}\phi\right)^2$ resulted to a
trivial non-minimal coupling function $\xi(\phi)$, however a
self-consistent framework emerged from the theory with correction
term $\sim \alpha' c \xi(\phi)\square \phi
\partial_{\mu}\phi \partial^{\mu}\phi$. The resulting
theory can easily be worked out analytically and we obtained an
inflationary theory that can easily be fitted with the Atacama
Cosmology Telescope constraints on the scalar spectral index and
the updated Planck constraints on the tensor-to-scalar ratio.
\end{abstract}


\maketitle

\section{Introduction}

The prominent candidate theory for early times, with observational
relevance to present day's physics, is inflation
\cite{inflation1,inflation2,inflation3,inflation4}. This theory
solves all the shortcomings of the hot Big Bang theory and will be
thoroughly tested by ground based Cosmic Microwave Background
(CMB) experiments like the Simons observatory
\cite{SimonsObservatory:2019qwx}, and from future gravitational
wave experiments
\cite{Hild:2010id,Baker:2019nia,Smith:2019wny,Crowder:2005nr,Smith:2016jqs,Seto:2001qf,Kawamura:2020pcg,Bull:2018lat,LISACosmologyWorkingGroup:2022jok}.
The Simons observatory and the future LiteBird experiment
\cite{LiteBIRD:2022cnt} will directly probe the B-mode in the CMB,
while the gravitational wave experiments will indirectly probe the
inflationary era by determining a stochastic gravitational wave
background compatible with some inflationary theory. So far the
Pulsar Timing Array experiments confirmed the existence of a
stochastic gravitational wave background in 2023
\cite{NANOGrav:2023gor,Antoniadis:2023ott,Reardon:2023gzh,Xu:2023wog},
but it is highly unlikely that inflation solely can explain this
signal \cite{Vagnozzi:2023lwo,Oikonomou:2023qfz}. The recent
Atacama Cosmology Telescope (ACT) data
\cite{ACT:2025fju,ACT:2025tim} combined with the DESI data
\cite{DESI:2024uvr}, stirred things up in inflationary physics,
since the scalar spectral index is in at least 2$\sigma$
discordance with the corresponding Planck data
\cite{Planck:2018jri}. To be specific, the scalar spectral index
is constrained by the ACT data to be,
\begin{equation}\label{act}
n_{\mathcal{S}}=0.9743 \pm
0.0034,\,\,\,\frac{\mathrm{d}n_{\mathcal{S}}}{\mathrm{d}\ln
k}=0.0062 \pm 0.0052\, .
\end{equation}
In addition, the updated Planck constraint on the tensor-to-scalar
ratio is \cite{BICEP:2021xfz},
\begin{equation}\label{planck}
r<0.036\, ,
\end{equation}
at $95\%$ confidence. There is already a large stream of articles
in the cosmology literature that aim to explain the ACT result
\cite{Kallosh:2025rni,Gao:2025onc,Liu:2025qca,Yogesh:2025wak,Yi:2025dms,Peng:2025bws,Yin:2025rrs,Byrnes:2025kit,Wolf:2025ecy,Aoki:2025wld,Gao:2025viy,Zahoor:2025nuq,Ferreira:2025lrd,Mohammadi:2025gbu,Choudhury:2025vso,Odintsov:2025wai,Q:2025ycf,Zhu:2025twm,Kouniatalis:2025orn,Hai:2025wvs,Dioguardi:2025vci,Yuennan:2025kde,Kuralkar:2025zxr,Kuralkar:2025hoz,Aoki:2025ywt},
although caution is needed for the moment in trusting the ACT
result \cite{Ferreira:2025lrd}. In this work we aim to examine
several scalar field theories with string origin, in view of the
ACT result. In string theory, the higher order corrections to the
low-energy effective action contain an infinite expansion with an
expansion parameter $\alpha'=\lambda_s^2$ where $\lambda_s$ is the
fundamental string scale. Restricting the action to the lowest
order gravitational action that ensures that the equations of
motion are second order, the first scalar field related
corrections to the scalar field action are $\sim \alpha'\xi(\phi)
\left( c \square \phi \partial_{\mu}\phi
\partial^{\mu}\phi +c_2\,\left(
\partial_{\mu}\phi \partial^{\mu}\phi\right)^2\right)$
\cite{Metsaev:1987zx,Cartier:2001is}. However, we will not attempt
any direct connection with a string theory framework, we will use
the fact that certain string theories produce the terms just
mentioned, that may appear in the low-energy scalar field
Lagrangian. However, no contact with an actual string theoretic
framework is attempted in this article. Specifically, in this work
we aim to present a self-consistent framework for studying
inflationary dynamics in such string corrected inflationary
theories. We shall develop the formalism that enables us to study
analytically such theories and the requirements for making such
analytic manipulations of the theory. As we shall show, the
resulting theory can be compatible with the ACT data for a wide
range of the free parameters of the theory, for approximately 60
$e$-foldings of inflation.

For this study, we shall assume that the spacetime is a flat
Friedmann-Robertson-Walker, with line element,
\begin{equation}
\label{FRW} \centering
ds^2=-dt^2+a(t)^2\sum_{i=1}^{3}{(dx^{i})^2}\, .
\end{equation}

\section{Self-consistent String Corrected Scalar Field Inflation Formalism}

In this section we shall develop the string corrected scalar field
inflationary theory aiming to present a self-consistent
theoretical framework. Before starting, let us note that the
inflationary Lagrangian of any sort will depend on the curvature
$R$, the scalar field and its derivatives denoted as
$X=\frac{1}{2}\partial_{\mu}\phi \partial^{\mu}\phi$ and the two
known fundamental constants of theoretical physics, the
cosmological constant $\Lambda$ and the reduced Planck mass $M_p$,
that is,
\begin{equation}\label{action1}
\mathcal{S}=\int \mathrm{d}^4x \mathcal{L}(R,X,\phi,M_p,\Lambda)\,
.
\end{equation}
Having this in mind, let us consider the following general
string-corrected scalar field action of the form,
\begin{equation}\label{actioncentral}
\mathcal{S}=\int \mathrm{d}^4x \sqrt{-g} \left(
\frac{R}{2\kappa^2}-\frac{1}{2}\partial_{\mu}\phi
\partial^{\mu}\phi-V(\phi)-c\xi(\phi)\square \phi \partial_{\mu}\phi
\partial^{\mu}\phi- c_2\xi(\phi)\left( \partial_{\mu}\phi
\partial^{\mu}\phi\right)^2\right)\, ,
\end{equation}
which contains solely higher derivatives of the scalar field  and
does not contain higher curvature corrections. In Eq.
(\ref{actioncentral}), $\kappa=\frac{1}{M_p}$ where $M_p$ is the
reduced Planck mass, and $c$ and $c_2$ are dimensionful constants,
which will depend on the fundamental constants of the theory, that
is $M_p$ and $\Lambda$ and $\xi(\phi)$ is a dimensionless function
of the scalar field. We shall consider separately the cases with
string corrections $\sim c\xi(\phi)\square \phi
\partial_{\mu}\phi \partial^{\mu}\phi$ and $\sim c_2\xi(\phi)\left( \partial_{\mu}\phi \partial^{\mu}\phi\right)^2$ for simplicity.

\subsection{Case I: String Corrections of the Form $\sim c\xi(\phi)\square \phi
\partial_{\mu}\phi \partial^{\mu}\phi$}

Let us consider the string corrected scalar field theory with the
following action,
\begin{equation}\label{actionstring1}
\mathcal{S}=\int \mathrm{d}^4x \sqrt{-g} \left(
\frac{R}{2\kappa^2}-\frac{1}{2}\partial_{\mu}\phi
\partial^{\mu}\phi-V(\phi)-c\xi(\phi)\square \phi \partial_{\mu}\phi
\partial^{\mu}\phi\right)\, ,
\end{equation}
which will prove the most interesting case phenomenologically and
also structurally since it provides a self-consistent theoretical
framework. Upon varying the gravitational action
(\ref{actionstring1}) with respect to the metric tensor and the
scalar field, we obtain the field equations,
\begin{equation}\label{field1}
\frac{3H^2}{\kappa^2}=\frac{\dot{\phi}^2}{2}+V+c\left(\dot{\xi}-6H\xi\right)\dot{\phi}^3\,
,
\end{equation}
\begin{equation}\label{field2}
-\frac{2\dot{H}}{\kappa^2}=\dot{\phi}^2+c\left(\dot{\xi}-6H\xi
\right)\dot{\phi}^3+c\dot{\phi}^2\left(2\xi
\ddot{\phi}\xi+\dot{\xi}\dot{\phi} \right)\, ,
\end{equation}
\begin{equation}\label{field3}
\ddot{\phi}+3H\dot{\phi}+V'+c\dot{\phi}\left(\ddot{\xi}\dot{\phi}+3\dot{\xi}\dot{\phi}-6\xi\left(\dot{H}\dot{\phi}+3\dot{\xi}\dot{\phi}-6\xi
\left(\dot{H}\dot{\phi}+2H\ddot{\phi}+3H^2\dot{\phi} \right)
\right) \right)=0\, .
\end{equation}
Now we will make a crucial assumption which can render the present
inflationary theory a simple theory to tackle analytically and
also a self-consistent theory. We assume that,
\begin{equation}\label{maincondition}
x=\frac{\dot{\xi}}{\xi H}\, ,
\end{equation}
where $x$ is an arbitrary dimensionless number the value of which
will be determined by demanding the self-consistency of the
theory. The motivation for the condition (\ref{maincondition})
comes from observing the field equations which can be simplified
if for example $\dot{\xi}=6 \xi H$. However we do not specify the
value of $x$ at this point, we leave it free to choose.

Due to the slow-roll conditions, it is natural to assume that the
potential in the Friedmann equation (\ref{field1}) overwhelms the
scalar field derivative terms, that is,
\begin{equation}\label{conditions1}
c\dot{\xi}\dot{\phi}^3\ll V,\,\,\,6H\xi\dot{\phi}^3\ll V\, ,
\end{equation}
and also we assume that the slow-roll conditions hold true,
\begin{equation}\label{conditions2}
|\dot{H}|\ll H^2\, .
\end{equation}
In view of the slow-roll approximations and the conditions
(\ref{maincondition}) and (\ref{conditions1}), the field equations
become,
\begin{equation}\label{field1a}
\frac{3H^2}{\kappa^2}\simeq V\, ,
\end{equation}
\begin{equation}\label{field2a}
-\frac{2\dot{H}}{\kappa^2}\simeq c(2x-6)H\xi \dot{\phi}^3\, ,
\end{equation}
\begin{equation}\label{field3a}
\dot{\phi}^2\simeq -\frac{V'}{c(x^2-18)H \xi}\, .
\end{equation}
where we also assumed that the slow-roll conditions extend to the
following quantities too,
\begin{equation}\label{conditions3}
\ddot{\phi}\ll\dot{\phi}^2,\,\,\, 3H\dot{\phi}\ll c \dot{\phi}^2
\xi H^2\, .
\end{equation}
As it will prove, the present theoretical framework is
self-consistent and the field equations produce one another for a
specific value of $x$. The hierarchy between the terms in the
field equations, is chosen as above, because the observation that
if $\dot{\xi}$ and $\xi H$ scale in the same way as functions of
time and produce a constant $x$, then the equations of motion are
rendered a closed system of equations, as we now show. So
basically, this is the motivation for assuming the dominance of
the corrections over the friction terms, and of course the
approximations must be checked in the end, if indeed they hold
true at the first horizon crossing. As we will show, the
approximations hold true for the values of the free parameters
that guarantee a viable phenomenology. Let us see this explicitly,
so starting from Eq. (\ref{field1a}) and taking the derivative, we
have,
\begin{equation}\label{aux1}
\frac{6\dot{H}H}{\kappa^2}=V'\dot{\phi}\, ,
\end{equation}
and upon substituting $\dot{H}$ from Eq. (\ref{field2a}) in Eq.
(\ref{aux1}), we obtain,
\begin{equation}\label{aux2}
\frac{6(x-3)}{x^2-18}=1\, ,
\end{equation}
which has two solutions $x=0$ and $x=6$. The solution $x=0$
results into a trivial theory, while $x=6$ is quite intriguing
because by looking Eq. (\ref{field1}) we can see that the field
equation can be simplified without any constraint. So from now on
we take $x=6$. Let us continue showing that the present
theoretical framework represented by the field equations
(\ref{field1a})-(\ref{field3a}) is self-consistent. So by starting
from Eq. (\ref{field1a}) and taking the derivative, we get Eq.
(\ref{aux1}). Now from Eq. (\ref{field2a}) due to Eq.
(\ref{field3a}) we have,
\begin{equation}\label{aux3}
-\frac{\dot{H}}{\kappa^2}=-c(x-3)H\xi
\dot{\phi}\frac{V'}{c(x^2-18)H^2\xi}\, ,
\end{equation}
so we have,
\begin{equation}\label{aux4}
\frac{6\dot{H}}{\kappa^2}=\frac{\dot{\phi}}{H}V'\, ,
\end{equation}
which is identical to Eq. (\ref{aux1}). So from Eq.
(\ref{field2a}) we ended up to Eq. (\ref{field1a}) using Eq.
(\ref{field3a}). Now from Eq. (\ref{field1a}) due to Eq.
(\ref{field3a}) we have,
\begin{equation}\label{aux5}
-\frac{\dot{H}}{\kappa^2}=\frac{c(x^2-18)}{6}\xi H \dot{\phi}^3\,
,
\end{equation}
which is identical with Eq. (\ref{field2a}) for $x=6$. Hence the
system of field equations (\ref{field1a})-(\ref{field3a}) is
closed to itself and self-consistent for $x=6$.

Now from the constraint Eq. (\ref{maincondition}), we get,
\begin{equation}\label{mainconditionaux}
\xi'\dot{\phi}=x\xi H\, ,
\end{equation}
so after some simple algebra, we get,
\begin{equation}\label{maindifferentialequation}
-\frac{(\xi')^2}{\xi^3}=\frac{c(x^2-18)x^2\kappa^4\,V^2}{9\,V'}\,
,
\end{equation}
which apparently constraints the scalar field potential $V(\phi)$
and the non-minimal coupling function $\xi(\phi)$, which cannot be
freely chosen, but must satisfy the differential equation
(\ref{maindifferentialequation}). Now in order to study the
inflationary phenomenology, one needs to find the slow-roll
indices and the $e$-foldings numbers expressed in terms of the
functions $\xi(\phi)$ and $V(\phi)$. The $e$-foldings number is
defined to be,
\begin{equation}\label{efoldings}
N=\int_{\phi_i}^{\phi_f}\frac{H}{\dot{\phi}}\mathrm{d}\phi\, ,
\end{equation}
and by using Eq. (\ref{mainconditionaux}) we have,
\begin{equation}\label{efoldingsfinal}
N=\frac{1}{x}\int_{\phi_i}^{\phi_f}\frac{\xi'}{\xi}\mathrm{d}\phi\,
,
\end{equation}
where $\phi_i$ is the value of the scalar field at the first
horizon crossing when inflation commences and $\phi_f$ is the
value of the scalar field at the end of the inflationary era. Now
for the string corrected theory at hand, the slow-roll indices are
\cite{Hwang:2005hb},
\begin{align}
\centering \label{indices} \epsilon_1&=-\frac{\dot
H}{H^2}&\epsilon_2&=\frac{\ddot\phi}{H\dot\phi}&\epsilon_3&=0&\epsilon_4&=\frac{\dot
E}{2HE}&\epsilon_5&=\frac{Q_a}{2HQ_t}&\epsilon_6&=\frac{\dot
Q_t}{2HQ_t}\, ,
\end{align}
with $Q_a=2c\xi \dot{\phi}^3 $, $Q_b=0$,
$E=\frac{1}{(\kappa\dot\phi)^2}\left(
\dot\phi^2+\frac{3Q_a^2}{2Q_t}+Q_c\right)$,
$Q_c=4c\dot{\phi}^3(x-3)\xi H$, $Q_d=-4c
\dot{\phi}^2(\dot{\xi}\dot{\phi}+\xi \ddot{\phi}-\xi\dot{\phi}H)$,
$Q_e=-8c\xi \dot{\phi}^3$, $Q_f=0$ and also
$Q_t=\frac{1}{\kappa^2}+\frac{Q_b}{2}$. Regarding the observable
quantities that determine the viability of the inflationary
regime, namely the scalar spectral index of the primordial
curvature perturbations and the tensor-to-scalar ratio, these are,
\begin{equation}\label{spectralindex}
n_{\mathcal{S}}=1+\frac{2 (-2
\epsilon_1-\epsilon_2-\epsilon_4)}{1-\epsilon_1}\, ,
\end{equation}
and also,
\begin{equation}\label{tensortoscalar}
r=\left |\frac{16 \left(c_A^3 \left(\epsilon_1-\frac{1}{4} \kappa
^2 \left(\frac{2
Q_c+Q_d}{H^2}-\frac{Q_e}{H}+Q_f\right)\right)\right)}{c_T^3
\left(\frac{\kappa ^2 Q_b}{2}+1\right)}\right |\, ,
\end{equation}
where $c_A$ is the sound speed of the scalar perturbations, with
its explicit form being,
\begin{equation}\label{soundspeed}
c_A=\sqrt{\frac{\frac{Q_a Q_e}{\frac{2}{\kappa ^2}+Q_b}+Q_f
\left(\frac{Q_a}{\frac{2}{\kappa
^2}+Q_b}\right)^2+Q_d}{\dot{\phi}^2+\frac{3 Q_a^2}{\frac{2}{\kappa
^2}+Q_b}+Q_c}+1}\, ,
\end{equation}
and moreover $c_T$ stands for the gravitational wave speed,
defined as,
\begin{equation}
\label{GW} \centering c_T^2=1-\frac{Q_f}{2Q_t}\, ,
\end{equation}
hence for the theory at hand $c_T^2=1$, hence no issues with the
GW170817 event emerge. This fact was expected, since the theory
contains higher derivatives of the scalar field and not of the
curvature. For this theory, the first slow-roll index acquires a
very simple form, which is,
\begin{equation}\label{epsilon1slowroll}
\epsilon_1=-\frac{3(x-3)x}{x^2-18}\frac{\xi}{\xi'}\frac{V'}{V}\, .
\end{equation}
In addition it is important to keep track of another quantity for
the viability of the inflationary era for this theoretical
framework, namely the amplitude of scalar perturbations
$\mathcal{P}_{\zeta}(k_*)$ and the constraints on it imposed by
the latest Planck data \cite{Planck:2018jri}. The definition of
the amplitude of the scalar perturbations is,
\begin{equation}\label{definitionofscalaramplitude}
\mathcal{P}_{\zeta}(k_*)=\frac{k_*^3}{2\pi^2}P_{\zeta}(k_*)\, ,
\end{equation}
which must be evaluated at the first horizon crossing, when the
inflationary era commenced, and $k_*$ stands for the CMB pivot
scale. The Planck data constrain the amplitude of the scalar
perturbations as follows
$\mathcal{P}_{\zeta}(k_*)=2.196^{+0.051}_{-0.06}\times 10^{-9}$
\cite{Planck:2018jri}, evaluated at the CMB pivot scale. The
amplitude of the scalar perturbations $\mathcal{P}_{\zeta}(k)$ can
be expressed in terms of the two point function $\zeta (k)$ of the
curvature perturbation, in the following way,
\begin{equation}\label{twopointfunctionforzeta}
\langle\zeta(k)\zeta (k')\rangle=(2\pi)^3 \delta^3 (k-k')
P_{\zeta}(k)\, .
\end{equation}
For the string corrected scalar field theory at hand, the
amplitude of the scalar perturbations $\mathcal{P}_{\zeta}(k)$ can
be expressed in terms of the slow-roll parameters in the following
way \cite{Hwang:2005hb},
\begin{equation}\label{powerspectrumscalaramplitude}
\mathcal{P}_{\zeta}(k)=\left(\frac{k \left((-2
\epsilon_1-\epsilon_2-\epsilon_4) \left(0.57\, +\log \left(\left|k
\eta \right| \right)-2+\log (2)\right)-\epsilon_1+1\right)}{(2 \pi
) \left(z c_A^{\frac{4-n_{\mathcal{S}}}{2}}\right)}\right)^2\, ,
\end{equation}
with $z=\frac{a \dot{\phi} \sqrt{\frac{E(\phi )}{\frac{1}{\kappa
^2}}}}{H (\epsilon_5+1)}$ and all the above quantities have to be
evaluated at the first horizon crossing, at which point, the
conformal time $\eta$ is equal to
$\eta=-\frac{1}{aH}\frac{1}{-\epsilon_1+1}$ \cite{Hwang:2005hb}.
Now having all the above relations at hand, one can easily examine
the inflationary phenomenology of the string corrected scalar
field theory at hand. The condition $|\epsilon_1|\sim
\mathcal{O}(1)$ will yield the value of the scalar field at the
end of inflation, and from Eq. (\ref{efoldingsfinal}) one can
evaluate the value of the scalar field at the beginning of
inflation. After that, substituting $\phi_i$ in all the above
quantities can give us a direct hint on whether the inflationary
theory is viable or not, bearing also in mind the constraints on
the amplitude of the scalar perturbations. We shall confront the
theory at hand with the ACT data (\ref{act}) and the updated
Planck constraints (\ref{planck}). In the next section we shall
give explicit examples of simple models that can generate a viable
string corrected scalar field inflationary theory. Before doing
that, in the next subsection we consider the theory with string
corrections $\sim c_2\xi(\phi)\left( \partial_{\mu}\phi
\partial^{\mu}\phi\right)^2$ and its consistency.

Before closing this section, let us discuss an interesting
question. One may raise the concern that the dominance of the
string-inspired correction terms over the canonical kinetic term
is implying that the effective-field-theory expansion breaks,
since the higher-order stringy corrections, would then be expected
to contribute at the same order as the kinetic term. This argument
however is not correct.

In general, the low-energy string effective action is basically an
expansion in terms of the string scale $\alpha'$, and it is not an
expansion in terms of the powers of the kinetic term $X \equiv
\dot{\phi}^{2}/2$. The fact that a specific $\mathcal{O}(\alpha')$
operator may be assumed to dominate the background dynamics does
not necessarily imply a loss of the effective field theory
control, given that higher-order operators remain perturbatively
suppressed by additional powers of the parameter $\alpha'$. In
particular, the effective field theory remains valid, as long as
the following hierarchy
\begin{equation}
H^{2} \ll M_{s}^{2} \equiv \alpha'^{-1}
\end{equation}
is satisfied, which is an assumption taken into account during the
whole the inflationary evolution considered in this work.

We need to note that it is well known that regimes in which
non-canonical kinetic terms may dominate over the canonical
contribution are perfectly consistent within effective field
theory frameworks. For examples of this sort, one may recall
$k$-inflation, DBI inflation, and Horndeski theories
\cite{Ohashi:2013pca,Arroja:2011yj,Taveras:2008yf,Armendariz-Picon:1999hyi,Garriga:1999vw,Choudhury:2012whm,Cai:2010wt,Langlois:2008qf,Chen:2005fe,Kobayashi:2019hrl,Lu:2020iav,Bettoni:2013diz},
in which the higher-derivative operators control the background
evolution, without however rendering the theory inconsistent. The
present model we discussed, belongs to this class of models.

Also, the background dominance of a higher-derivative operator
does not directly implies that higher-order
$\mathcal{O}(\alpha'^2)$ corrections can become comparable. Such
terms are in general suppressed by additional factors of the
parameter $\alpha' H^{2}$ and thus remain subleading as long as
$H^{2}/M_{s}^{2} \ll 1$. The slow-roll conditions we imposed
define a general hierarchy among background contributions, not a
hierarchy in the effective field theory expansion parameter.

We also stress that the consistency of the effective field theory
is mainly governed by the behavior of cosmological perturbations,
and not solely by the background equations. In the present
framework, the quadratic action for the scalar perturbations is
well defined, and the scalar sound speed remains finite, and we
have no strong-coupling inconsistencies. This confirms that the
inflationary solutions are valid, despite the dominance of the
leading string correction in the background dynamics.

Finally, let us also mention that the argument that corrections
should not overwhelm canonical terms is not valid in other
theories too, for example in $R^2$ theory, the $R^2$ term
overwhelms the Einstein Hilbert term $R$, although the $R^2$ term
is a correction. One may think these in the context of corrections
of the standard canonical scalar field action, in which the $R^2$
term may dominate the leading order kinetic terms of the scalar
field.

We need to note that the expressions for the spectral index, the
tensor-to-scalar ratio and other observables, hold true if the
cosmological perturbations freeze on superhorizon scales. This was
proven that it indeed holds true in Ref. \cite{Hwang:2005hb} and
we quote here the proof, following closely \cite{Hwang:2005hb},
see \cite{Hwang:2005hb} for more details. We perturb the flat FRW
metric as follows,
 \bea
   & & d s^2 = - a^2 \left( 1 + 2 \alpha \right) d \eta^2
       - 2 a^2 \beta_{,\alpha} d \eta d x^\alpha
       + a^2 \left( g^{(3)}_{\alpha\beta}
       + 2 \varphi g^{(3)}_{\alpha\beta}
       + 2 \gamma_{,\alpha|\beta}
       + 2 C_{\alpha\beta} \right) d x^\alpha d x^\beta,
   \label{metric}
\eea with $a(\eta)$ being the cosmic scale factor in terms of the
conformal time $\eta$. Here $\alpha$, $\beta$, $\gamma$ and
$\varphi$ are scalar spacetime-dependent perturbations, and the
tensor perturbation $C_{\alpha\beta}$ is a transverse and a
trace-free tensor. The metric $g^{(3)}_{\alpha\beta}$ denotes the
comoving three-space section of the FRW metric, \bea
   g^{(3)}_{\alpha\beta} d x^\alpha d x^\beta
   &=& {1 \over \left( 1 + \bar r^2 \right)^2}
       \left( d x^2 + d y^2 + d z^2 \right)\,. \eea
The kinematic quantities expressed in the normal frame are as
follows, \bea
   & & \theta = 3 H , \quad
       \sigma_{\alpha\beta}
       = \chi_{,\alpha|\beta}
       - {1 \over 3} g_{\alpha\beta}^{(3)} \Delta \chi
       + a^2 \dot C^{(t)}_{\alpha\beta}, \quad
       a_\alpha = \alpha_{,\alpha}, \quad
       R^{(h)} = {1 \over a^2} \left[ 6 K - 4 \Delta
       \varphi \right],
   \label{kinematic-quantities}
\eea with \bea
   & & \chi \equiv a \left( \beta + a \dot \gamma \right),
   \label{chi-def}
\eea and $\Delta$ is the Laplacian operator in
$g^{(3)}_{\alpha\beta}$ and also $\theta$ stands for the expansion
scalar, $\sigma_{ab}$ stands for the shear tensor, and $a_a$
denotes the acceleration vector. We use the following gauge
transformation, $\hat x^a \equiv x^a + \tilde \xi^a (x^e)$ and we
have the expressions for the perturbation variables, \bea
   & & \hat \alpha = \alpha - \dot \xi^t, \quad
       \hat \beta = \beta - {1 \over a} \xi^t
       + a \left( {\xi \over a} \right)^\cdot, \quad
       \hat \gamma = \gamma - {1 \over a} \xi, \quad
       \hat \varphi = \varphi - H \xi^t, \quad
       \hat \chi = \chi - \xi^t, \quad
       \hat \kappa = \kappa
       + \left( 3 \dot H + {\Delta \over a^2} \right) \xi^t,
   \nonumber \\
   & & \delta \hat \mu = \delta \mu - \dot \mu \xi^t, \quad
       \delta \hat p = \delta p - \dot p \xi^t, \quad
       \hat v = v - {1 \over a} \xi^t, \quad
       \hat \Pi = \Pi, \quad
       \delta \hat \phi = \delta \phi - \dot \phi \xi^t; \quad
       \hat C_{\alpha\beta} = C_{\alpha\beta}, \quad
       \hat \Pi_{\alpha\beta}^{({t})} = \Pi_{\alpha\beta}^{({t})},
   \label{GT}
\eea where $\xi^0 \equiv {1 \over a} \xi^t$ and in addition
$\xi_\alpha \equiv \xi_{,\alpha}$. Furthermore, $\bar \phi$ and
$\delta \phi$ denote the background and the perturbation part of
the scalar field $\phi({\bf x}, t)$. We use the gauge-invariant
variables, \bea
   & & \varphi_\chi \equiv \varphi - H \chi, \quad
       \varphi_v \equiv \varphi - a H v, \quad
       \delta_v \equiv \delta - a {\dot \mu \over \mu} v, \quad
       \delta \phi_\varphi \equiv \delta \phi
       - {\dot \phi \over H} \varphi
       \equiv - {\dot \phi \over H} \varphi_{\delta \phi}, \quad
       v_\chi \equiv v - {1 \over a} \chi
       \equiv - {1 \over a} \chi_v,
   \label{GI-variables}
\eea where $\delta \equiv \delta \mu / \mu$. Also we use the
following perturbation variable definitions, \bea
   & & \Phi \equiv \varphi_{\delta \phi}, \quad
       \Psi \equiv \varphi_\chi + {\dot F + Q_a \over 2F + Q_b}
       {\delta F_\chi \over \dot F}.
   \label{Psi-def-string}
\eea and also, \bea
   & & c_A^2 = 1 + { Q_d + {\dot F + Q_a \over 2F + Q_b} Q_e
       + \left( {\dot F + Q_a \over 2F + Q_b} \right)^2 Q_f
       \over \omega \dot \phi^2 + 3 {(\dot F + Q_a)^2 \over 2 F + Q_b} + Q_c }.
\eea The scalar-type perturbation equations considered in
\cite{Hwang:2005hb} for higher derivative theories,  can be
written as follows, \bea
   & & \dot \Phi = 2 x_1 {\Delta \over a^2} \Psi,
   \label{dot-Phi-eq} \\
   & & {1 \over x_2} \left( x_2 \Psi \right)^\cdot = {1 \over 2} x_3
   \Phi,
   \label{dot-Psi-eq}
\eea where,
\begin{equation}\label{x1}
x_1={ \left( H + {\dot F + Q_a \over 2 F + Q_b} \right)
         \left( F + {1 \over 2} Q_b \right)
         \over \omega \dot \phi^2
         + 3 {(\dot F + Q_a)^2 \over 2 F + Q_b} + Q_c }\, ,
\end{equation}
\begin{equation}\label{x2}
 x_2={a (F + {1 \over 2} Q_b) \over
       H + {\dot F + Q_a \over 2 F + Q_b}}\, ,
\end{equation}
\begin{equation}\label{x3}
x_3={1 \over \left( H + {\dot F + Q_a \over 2 F + Q_b} \right)
         \left( F + {1 \over 2} Q_b \right)} x_4\, ,
\end{equation}
\begin{equation}\label{x24}
x_4=\dot \phi^2 + 3 {(\dot F + Q_a)^2 \over 2 F + Q_b}
  + Q_c + Q_d + {\dot F + Q_a \over 2F + Q_b} Q_e
  + \left( {\dot F + Q_a \over 2F + Q_b} \right)^2 Q_f\, ,
\end{equation}
and $F=\frac{1}{\kappa^2}$. After some normalization and
redefinitions of variables, as follows, \bea
   & & z= \frac{a\dot{\phi}}{H}, \quad
    \quad
       \tilde v \equiv z \Phi, \quad
       u \equiv \frac{1}{\kappa^2}\frac{a}{H} {1 \over z} \Psi,
   \label{u-v-def}
\eea we end up to the perturbation equations,
 \bea
   & & \tilde v^{\prime\prime}
       - \left( c_A^2 \Delta + {z^{\prime\prime} \over z} \right) \tilde v
       = a^2 z \left[ {1 \over a z^2} \left( a z^2 \dot \Phi \right)^\cdot
       - c_A^2 {\Delta \over a^2} \Phi \right] = 0,
   \label{v-eq} \\
   & & u^{\prime\prime} - \left[ c_A^2 \Delta
       + {(1/\bar z)^{\prime\prime} \over (1/\bar z)} \right] u
       = {a^2 x_2 \over \bar z}
       \left\{ {\bar z^2 \over a x_2} \left[ {a \over \bar z^2}
       \left( x_2 \Psi \right)^\cdot \right]^\cdot
       - c_A^2 {\Delta \over a^2} \Psi \right\} = 0.
   \label{u-eq}
\eea In these differential equations of the perturbations, $c_A$
plays the role of the wave speed of the fluctuating fluid or field
and the simultaneously excited metric.

For the tensor mode, using the following \bea
   & & z_t \equiv a \sqrt{Q_t}, \quad
       v_t \equiv z_t \Phi,
   \label{z-v-def-GW}
\eea with $\Phi = C_{\alpha\beta}$ or $h_{\ell {\bf k}}$, we get
\bea
   & & v_t^{\prime\prime}
       - \left( c_T^2 \Delta + {z_t^{\prime\prime} \over z_t} \right) v_t
       = a^2 z_t \left[
       {1 \over a z_t^2} \left( a z_t^2 \dot \Phi \right)^\cdot
       - c_T^2 {\Delta \over a^2} \Phi \right]
       = 0.
   \label{v-eq-GW}
\eea The differential equations (\ref{v-eq},\ref{v-eq-GW}) are
generally valid for a wide variety of higher derivative gravity
theories. In the large-scale limits (superhorizon scales), with
$c_A^2 k^2 \ll z^{\prime\prime}/z$ and $(1/\bar
z)^{\prime\prime}/(1/\bar z)$, we get \bea
   & & \Phi (k, \eta)
       = {1 \over z} \tilde v
       = C (k) \Bigg\{ 1
       + k^2 \left[ \int^\eta \bar z^2
       \left( \int^\eta {d \eta \over z^2} \right) d \eta
       - \int^\eta \bar z^2 d \eta \int^\eta {d \eta \over z^2} \right] \Bigg\}
       - 2 \tilde d (k) k^2 \int^\eta {d \eta \over z^2},
   \\
   & & \Psi (k, \eta)
       = {\bar z \over x_2} u
       = C (k) {1 \over 2 x_2} \int^\eta \bar z^2 d \eta
       + \tilde d (k) {1 \over x_2} \Bigg\{ 1
       + k^2 \left[ \int^\eta {1 \over z^2}
       \left( \int^\eta \bar z^2 d \eta \right) d \eta
       - \int^\eta \bar z^2 d \eta \int^\eta {d \eta \over z^2} \right]
       \Bigg\}.\eea
Notice that at leading order in the large-scale expansion, the
$C$-mode of the perturbation $\Phi$ remains constant on
superhorizon scales. Thus, by ignoring the transient mode, we get
\bea
   & & \Phi (k, \eta) = C (k).
\eea and it is remarkable to point out that it remains constant
regardless the equation of state $p(\mu)$ (of the matter fluid
content, with pressure $p$ and energy density $\mu$), the field
potential $V(\phi)$, and the gravity theories $f(\phi,R,X)$,
$\omega(\phi)$, $\xi(\phi)$.

\subsection{Case II: String Corrections of the Form $\sim c_2\xi(\phi) \left(\partial_{\mu}\phi \partial^{\mu}\phi \right)^2$}

Let us now consider another class of string corrected scalar field
theory with the following action,
\begin{equation}\label{actionstring1add}
\mathcal{S}=\int \mathrm{d}^4x \sqrt{-g} \left(
\frac{R}{2\kappa^2}-\frac{1}{2}\partial_{\mu}\phi
\partial^{\mu}\phi-V(\phi)-c_2\xi(\phi) \left(\partial_{\mu}\phi \partial^{\mu}\phi \right)^2\right)\, ,
\end{equation}
which will prove not so interesting phenomenologically. Upon
varying the gravitational action (\ref{actionstring1}) with
respect to the metric and the scalar field, we obtain the field
equations,
\begin{equation}\label{field1add}
\frac{3H^2}{\kappa^2}=\frac{\dot{\phi}^2}{2}+V-c_2\xi\dot{\phi}^4\,
,
\end{equation}
\begin{equation}\label{field2add}
-\frac{2\dot{H}}{\kappa^2}=-2c_2\xi \dot{\phi}^4\, ,
\end{equation}
\begin{equation}\label{field3add}
\ddot{\phi}+3H\dot{\phi}+V'+c_2\dot{\phi}^2\left(-3\dot{x}\dot{\phi}-12\xi
\ddot{\phi}-12 H\xi \dot{\phi} \right)=0\, .
\end{equation}
Now we will make again the assumption (\ref{maincondition}) and
also we assume that the slow-roll conditions hold true, so the
field equations become,
\begin{equation}\label{field1aadd}
\frac{3H^2}{\kappa^2}\simeq V\, ,
\end{equation}
\begin{equation}\label{field2aadd}
\frac{\dot{H}}{\kappa^2}\simeq c_2\xi \dot{\phi}^4\, ,
\end{equation}
\begin{equation}\label{field3aadd}
\dot{\phi}^3\simeq \frac{2V'}{3(x+4)\,c_2 \xi H}\, .
\end{equation}
Now let us see whether the above set of field equations is
self-consistent and closed to itself. From Eq. (\ref{field1aadd})
by taking the derivative we obtain,
\begin{equation}\label{aux1add}
\frac{6\dot{H}H}{\kappa^2}=V'\dot{\phi}\, ,
\end{equation}
and from Eq. (\ref{field2aadd}), we obtain,
\begin{equation}\label{aux2}
\frac{x+4}{4}=1\, ,
\end{equation}
which has the solutions $x=0$, however this is a trivial solution
since it yields $\xi=\mathrm{constant}$. Thus we shall not further
analyze this class of string corrected scalar field theories. We
need to note that the choice $x=$constant is made after the field
equations are derived and this is a restriction on the dynamics of
inflation, but does not restrict the dynamics before the variation
of the action. In some sense it is a restriction on some subspace
of the total phase space of the system. In addition, we assumed
that $x=$constant, in the scenario that $x$ is not constant, the
whole analysis collapses because the solution or the trajectories
in the total phase space, actual propagate through the entire
phase space. Furthermore, the system of equations would not be a
closed system exactly reproducible. In addition, we tried the
$x\neq \mathrm{constant}$ case, but it is truly impossible to
derive any analytic results. There are too many terms that mix in
a non-hierarchical or manageable way, hence it is too difficult to
determine the dynamics analytically, even in the slow-roll case.

\section{String Corrected Scalar Field Inflation Models Compatible with the ACT Data}

Let us now present some viable scenarios of string corrected
scalar field theory, which are compatible with both the ACT data
and the updated Planck constraints on the tensor-to-scalar ratio.

\begin{figure}
\centering
\includegraphics[width=30pc]{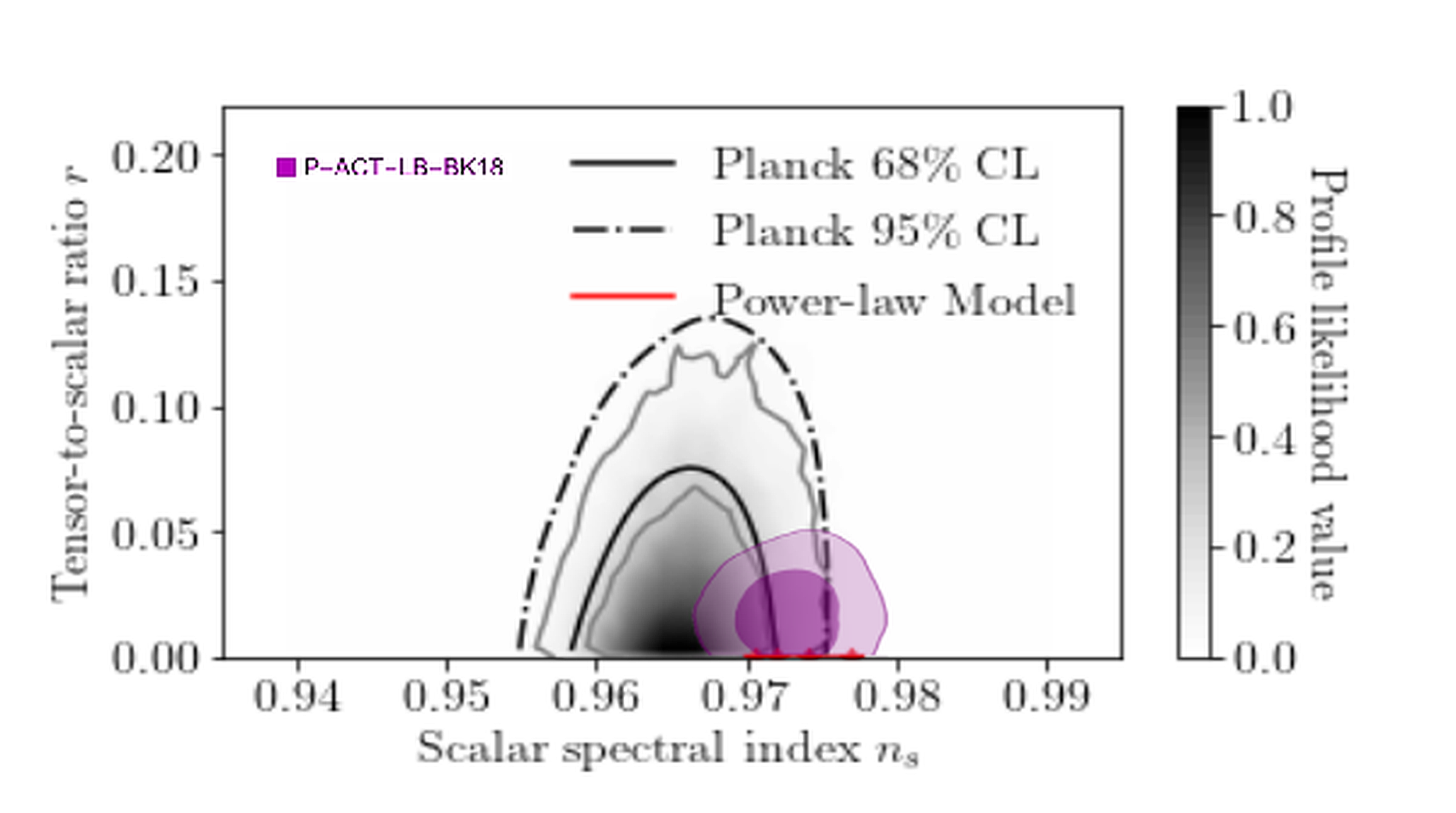}
\caption{Marginalized curves of the Planck 2018 data and power-law
model (\ref{xi11}) versus the ACT data, the Planck 2018 data and
the updated Planck tensor-to-scalar ratio
constraint.}\label{plot2}
\end{figure}

We start off with a power-law scenario for the scalar coupling
function $\xi(\phi)$, in which case the function $\xi(\phi)$ is
chosen to be,
\begin{equation}\label{xi11}
\xi(\phi)=\frac{(b+\lambda  (\kappa  \phi ))^n}{M}\, ,
\end{equation}
where $\lambda=\frac{\delta M_p^2}{\Lambda}$, $b$ is a
dimensionless parameter and $M=\frac{\Lambda ^3 \mu }{M_p^6}$ and
$\mu$ is a dimensionless parameter. By solving the differential
equation (\ref{maindifferentialequation}), given $\xi(\phi)$ in
Eq. (\ref{xi11}) we obtain the scalar potential $V(\phi)$,
\begin{equation}\label{V11}
V(\phi)=\frac{\kappa ^2 \lambda ^2 M (n+3)}{\frac{72 c \kappa ^3
(b+\kappa  \lambda  \phi )^{n+3}}{\lambda n^2}+c_1}\, ,
\end{equation}
where $c_1$ is a dimension $[m]^{-4}$ integration constant of the
form $c_1=\frac{\beta}{\Lambda^{-2}}$ and $\beta$ is
dimensionless. For the given scalar potential and coupling
function $\xi (\phi)$, the first slow-roll index $\epsilon_1$ from
Eq. (\ref{epsilon1slowroll}) reads in this case,
\begin{equation}\label{epsilon1model11}
\epsilon_1=\frac{216 c \kappa ^3 (n+3) (b+\kappa  \lambda \phi
)^{n+3}}{n \left(72 c \kappa ^3 (b+\kappa  \lambda \phi
)^{n+3}+c_1 \lambda  n^2\right)}\, , \, ,
\end{equation}
so upon solving $\epsilon_1(\phi_f)=1$ we obtain,
\begin{equation}\label{phifmodel11}
\phi_f=\frac{\left(\frac{c_1 \lambda  n^3}{c \kappa ^3 (144
n+648)}\right)^{\frac{1}{n+3}}-b}{\kappa  \lambda }\, .
\end{equation}
Upon plugging this in the expression for the $e$-foldings number
in Eq. (\ref{efoldingsfinal}), given $\xi(\phi)$, we get $\phi_i$,
which is for this model,
\begin{equation}\label{phiimodel11}
\phi_i=\frac{e^{-\frac{6N}{n}} \left(\frac{c_1 \lambda n^3}{c
\kappa ^3 (144 n+648)}\right)^{\frac{1}{n+3}}-b}{\kappa \lambda
}\, .
\end{equation}
Now one can obtain a viable phenomenology compatible with the ACT
data (\ref{act}) and the updated Planck constraints
(\ref{planck}), by choosing in this case for example
$\delta=9.08099\times 10^{-38}$ and $\beta=4.55945 \times 10^{71}$
for $N=60$ and $b=2$, in which case the scalar spectral index
becomes $n_{\mathcal{S}}=0.97428$ and the tensor-to-scalar ratio
becomes $r=7.89\times 10^{-174}$, practically zero, and the
amplitude of the scalar perturbations becomes
$\mathcal{P}_{\zeta}(k_*)=2.183\times 10^{-9}$. In Fig.
\ref{plot2} we confront the model at hand with the ACT and Planck
2018 likelihood curves and the updated Planck 2018 constraints on
the tensor-to-scalar ratio for $\beta$ chosen in the range
$\beta=[6.839\times 10^{70}, 2.097\times 10^{72}]$.
As it can be seen in Fig. \ref{plot2}, the model (\ref{xi11}) is
well fitted inside the ACT data (\ref{act}) and the Planck
constraint (\ref{planck}).

\begin{figure}
\centering
\includegraphics[width=30pc]{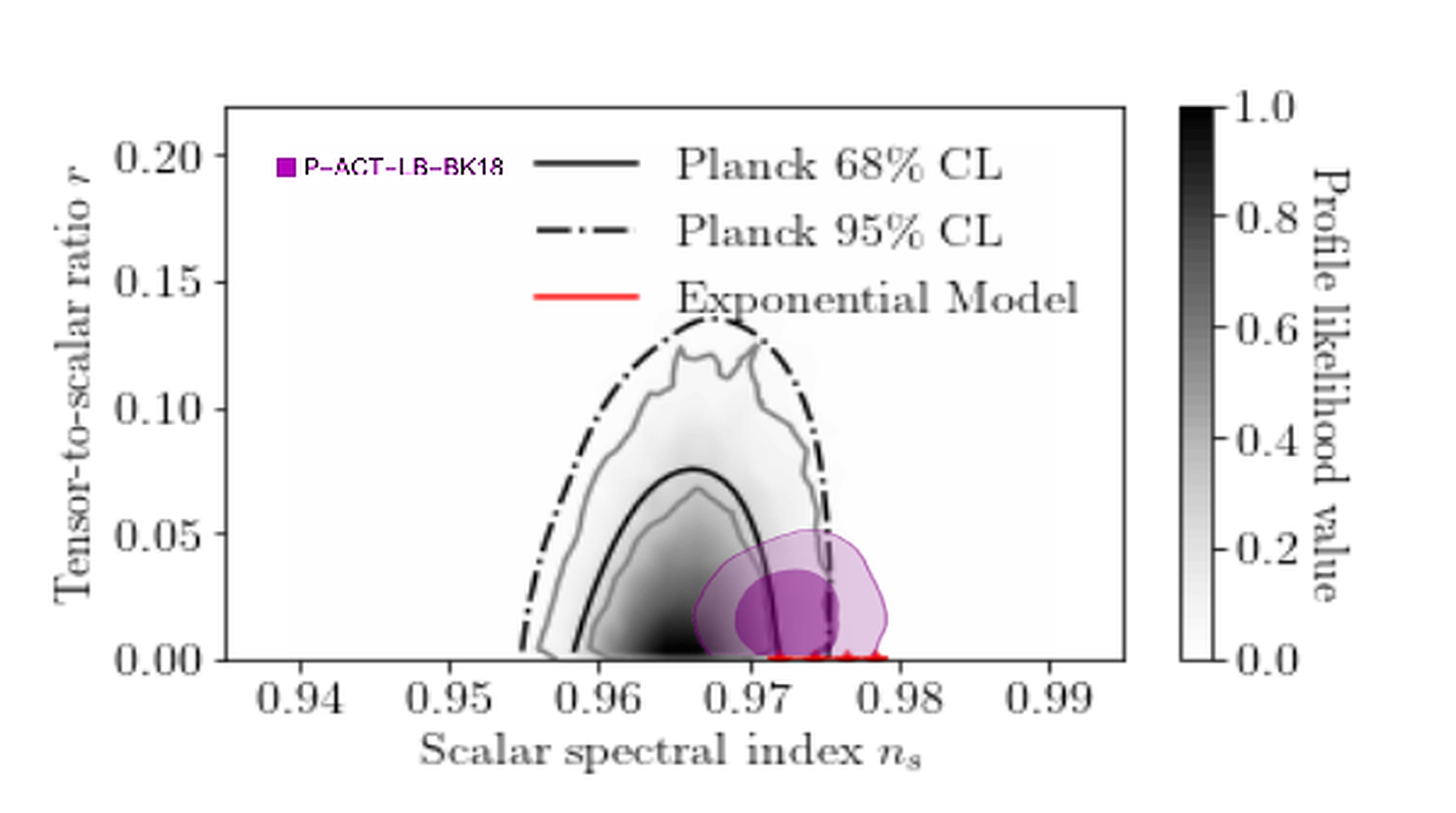}
\caption{Marginalized curves of the Planck 2018 data and power-law
model (\ref{xi111}) versus the ACT data, the Planck 2018 data and
the updated Planck tensor-to-scalar ratio
constraint.}\label{plot3}
\end{figure}

Let us now consider an exponential model in which case we choose
the coupling function $\xi(\phi)$ as follows,
\begin{equation}\label{xi111}
\xi(\phi)=\lambda  \exp (\gamma  (\kappa  \phi ))\, ,
\end{equation}
where $\lambda$ is a dimensionless constant and
$\gamma=\frac{\delta M_p^2}{\Lambda}$. Again by solving the
differential equation (\ref{maindifferentialequation}), given
$\xi(\phi)$ in Eq. (\ref{xi111}) we obtain the scalar potential
$V(\phi)$,
\begin{equation}\label{V111}
V(\phi)=-\frac{\gamma ^3}{\gamma ^3 c_1-72 c \kappa \lambda
e^{\gamma  \kappa  \phi }}\, ,
\end{equation}
where $c_1$ is again a dimension $[m]^{-4}$ integration constant
of the form $c_1=\frac{\beta M_p^3}{\Lambda^{3/2}}$. For the given
scalar potential and coupling function $\xi (\phi)$, the first
slow-roll index $\epsilon_1$ from Eq. (\ref{epsilon1slowroll})
reads in this case,
\begin{equation}\label{epsilon1model111}
\epsilon_1=-\frac{216 c \kappa  \lambda  e^{\gamma  \kappa \phi
}}{\gamma ^3 c_1-72 c \kappa  \lambda e^{\gamma \kappa  \phi }}\,
,
\end{equation}
so in this case, upon solving $\epsilon_1(\phi_f)=1$ we obtain,
\begin{equation}\label{phifmodel111}
\phi_f=\frac{\log \left(-\frac{\gamma ^3 c_1}{144 c \kappa \lambda
}\right)}{\gamma  \kappa }\, .
\end{equation}
Upon plugging this in the expression for the $e$-foldings number
in Eq. (\ref{efoldingsfinal}), given $\xi(\phi)$, we get $\phi_i$,
which is for this model,
\begin{equation}\label{phiimodel111}
\phi_i=\frac{\log \left(-\frac{\gamma ^3 c_1}{144 c \kappa \lambda
}\right)-6 N}{\gamma  \kappa }\, .
\end{equation}
Now one can obtain a viable phenomenology compatible with the ACT
data (\ref{act}) and the updated Planck constraints
(\ref{planck}), by choosing in this case for example
$\delta=1.2557\times 10^{-42}$, $\beta=-7.585 \times 10^{-20}$,
for $N=60$ and $\lambda=10^{1.9}$ and $c=10^{-3.2}$, in which case
the scalar spectral index becomes $n_{\mathcal{S}}=0.974$ and the
tensor-to-scalar ratio becomes $r=4.0952 \times 10^{-171}$ and in
addition, the amplitude of the scalar perturbations becomes
$\mathcal{P}_{\zeta}(k_*)=2.198\times 10^{-9}$. In Fig.
\ref{plot3} we confront the model at hand with the ACT and Planck
2018 likelihood curves and the updated Planck 2018 constraints on
the tensor-to-scalar ratio for $\delta$ in the range
$\delta=[1.141\times 10^{-42},1.304\times 10^{-42}]$.
As it can be seen in Fig. \ref{plot3}, the model (\ref{xi111}) is
well fitted inside the ACT data (\ref{act}) and the Planck
constraint (\ref{planck}).

Thus in this section we demonstrated that the string corrected
scalar field theory framework can provide viable inflationary
phenomenology compatible with both the ACT data and the updated
Planck constraints. The characteristic of the two models we
presented is the almost zero tensor-to-scalar ratio, but this has
to be a model-dependent feature.

Before closing, let us briefly discuss the approximations we made
in the field equations, and if these are justified in the end, for
the values of the free parameters we chose for the exponential and
the power-law model. In the first place, a justifiable objection
is that the kinetic term $\frac{\dot{\phi}^2}{2}$ cannot be
subleading compared to $c(2x-6)H\xi \dot{\phi}^3$ which originates
from a string correction origin. This might be indeed logical,
however using this assumption, we ended up to a closed set of
field equations, which is self-consistent and can become
ACT-compatible. Moreover, the approximation
$\frac{\dot{\phi}^2}{2}\ll c(2x-6)H\xi \dot{\phi}^3$ is found
numerically that it holds true, for example for the power-law
model, using the set of values of the free parameters for which
the ACT-compatibility is guaranteed we ended up to
$\frac{\dot{\phi}^2}{2}=10^{-26}M$ and $c(2x-6)H\xi
\dot{\phi}^3=10^{-20}M$ in Planck units. Also for the exponential
model we get $\frac{\dot{\phi}^2}{2}=10^{-157}/c_1$ and
$c(2x-6)H\xi \dot{\phi}^3=110^{-155}/c_1$, which validates the
assumption.

However we need to note that the results indicate that a certain
level of fine-tuning of the variables is required in order to
obtain phenomenologically viable results. This feature counts on
the downside of this theoretical framework.

In addition, since the theory studied in this work is a potential
driven scalar field theory, one expects that the reheating era
will be initiated by small oscillations of the scalar field. We
will not go into details since the full treatment would require
careful analysis, but the theory is a scalar theory and we expect
that the field oscillations will drive the reheating stage.

\section{Conclusions}

In this work we considered the first low-energy string corrections
to the single scalar field inflationary Lagrangian. Specifically,
it is known that in string theory the higher order corrections to
the low-energy effective action contain an infinite expansion with
an expansion parameter $\alpha'=\lambda_s^2$ with $\lambda_s$
being the fundamental string scale. If one restricts the action to
the lowest order gravitational action, which ensures that the
equations of motion are second order, the first corrections to the
single scalar field action are $\sim \alpha'\xi(\phi) \left( c
\square \phi
\partial_{\mu}\phi
\partial^{\mu}\phi +c_2\,\left(
\partial_{\mu}\phi \partial^{\mu}\phi\right)^2\right)$
\cite{Metsaev:1987zx,Cartier:2001is}. We examined the effect of
these two correction terms independently on the inflationary
dynamics of single scalar field inflation. Using only the
slow-roll assumption, we aimed to develop a formalism that is
self-consistent, so the approximations used were not just leading
order corrections to field equations, but the field equations
reproduce one the other. The theory with correction term $\sim
\alpha'c_2\,\left(
\partial_{\mu}\phi \partial^{\mu}\phi\right)^2$ did not result to
interesting results but yielded a trivial inflationary theory with
constant non-minimal coupling function $\xi(\phi)$, which we did
not further study. However, the theory with the correction term
$\sim \alpha'\left( c \square \phi
\partial_{\mu}\phi \partial^{\mu}\phi\right)$ yielded quite
interesting results, since the resulting theory was found
self-consistent, with the field equations reproducing one another.
We found analytically the slow-roll indices and the $e$-foldings
number in closed forms and we examined two models, with the
non-minimal coupling function $\xi(\phi)$ having a power-law form
and an exponential form. Accordingly, the scalar potential was
found via the differential equation that it is required to
satisfy, which also contains the non-minimal coupling function
$\xi(\phi)$. The models were confronted with both the ACT data and
the updated Planck constraints on the tensor-to-scalar ratio and
the results indicated that the models can easily be compatible
with the ACT data, for 60 $e$-foldings. One non-trivial extension
of this work is to try to develop a theory that simultaneously
considers the above string corrections, or even including
Einstein-Gauss-Bonnet terms. This task however exceeds the
purposes of this letter. Finally, it is interesting to note that
the class of models we studied in this work may generate
interesting dark energy phenomena, such as phantom divide
crossings \cite{Tsujikawa:2025wca}. Also it is interesting to note
that small tensor-to-scalar ratios can also be achieved by
scalar-tensor theories \cite{Kouniatalis:2025orn}. We further need
to note that the ACT data assume overall base $\Lambda$CDM,
however the DESI data indicated that the dark energy might be
dynamical \cite{Capozziello:2025qmh}. Such issues must be taken
into account when the ACT inflationary data are considered, as for
example in \cite{Yuennan:2026fcn}. One thing is certain, it is too
early to make a decisive conclusion about the tilt difference in
the spectral index between Planck and ACT. There are strong hints
though. This debatable difference also propagates on the running
of the spectral index, at $1-\sigma$ significance. The ACT data
point out a positive spectral index. This is sensational, so the
future data are eagerly anticipated. In addition, it is important
to stress that the value of the spectral index is inferred under
the assumption that the $\Lambda$CDM model holds
pre-recombination, and that any type of new physics introduced
pre-recombination for example in order to solve the Hubble
tension, (such as early dark energy), typically shifts the
spectral index upwards towards the Harrison-Zeldovich scale
invariant value  usually to compensate for changes in the damping
scale.

\end{document}